\documentclass[runningheads]{llncs}

	\title{Querying collections of tree-structured records  in the presence of within-record referential constraints}

\titlerunning{ } 

\author{Foto N. Afrati \and
	Matthew Damigos}
\authorrunning{F. Afrati and M. Damigos}

\institute{National Technical University of Athens\\
	\email{\{afrati,mgdamig\}@gmail.com}}

\usepackage[ruled,vlined]{algorithm2e}
\usepackage{listings}
\usepackage{xcolor}
\usepackage{graphicx}
\usepackage{hyperref}
\usepackage{tikz}
\usepackage{amsmath}
\usepackage{fancyvrb}

\usepackage{caption}
\usepackage[subrefformat=parens,labelformat=parens]{subcaption} 
\usepackage{tabularx} 
\usepackage{multirow}

\usepackage{caption}
\usepackage[subrefformat=parens,labelformat=parens]{subcaption} 
\usepackage{tabularx} 
\usepackage{multirow}

\usetikzlibrary{arrows,positioning,shapes,backgrounds,fit}
\tikzstyle{every picture}+=[remember picture]
\tikzstyle{format} = [draw, thin, fill=blue!20]
\tikzstyle{medium} = [ellipse, draw, thin, fill=green!20, minimum height=2.5em]
\tikzstyle{punkt} = [rectangle, rounded corners, shade, top color=white, bottom color=blue!50!black!20, draw=blue!40!black!60, very thick]
\tikzstyle{arr} = [single arrow, draw, rotate=270, fill=red!50]

\newcommand{\CD}{{\mathcal D}} 
\newcommand{\CC}{{\mathcal C}} 
\newcommand{\CL}{{\mathcal L}} 
\newcommand{\CT}{{\mathcal T}} 
\newcommand{\idarr}{\xrightarrow{\scriptsize 1}} 
\newcommand{\refarr}{\xrightarrow{r}} 

\newcommand{\squishlist}{
	\begin{list}{$\bullet$}
		{ \setlength{\itemsep}{0pt} \setlength{\parsep}{3pt}
			\setlength{\topsep}{3pt} \setlength{\partopsep}{0pt}
			\setlength{\leftmargin}{1.5em} \setlength{\labelwidth}{1em}
			\setlength{\labelsep}{0.5em} } }
		
	\newcommand{\squishlisttwo}{
		\begin{list}{$\bullet$}
			{ \setlength{\itemsep}{0pt} \setlength{\parsep}{0pt}
				\setlength{\topsep}{0pt} \setlength{\partopsep}{0pt}
				\setlength{\leftmargin}{2em} \setlength{\labelwidth}{1.5em}
				\setlength{\labelsep}{0.5em} } }
		\newcommand{\squishend}{	
	\end{list}  }

\begin{document}

\maketitle

\begin{abstract}
In this paper, we consider a tree-structured data model used in many commercial databases like Dremel, F1, JSON stores.
We define identity and referential constraints within each tree-structured record. 
The query language is a variant of SQL and flattening is used as an evaluation mechanism.
We investigate querying in the presence of these constraints, and point out the challenges that arise from taking them into account during query evaluation.
\end{abstract}

	\section{Introduction}

Systems that efficiently analyze  complex data (e.g., graph, or hierarchical data) are ubiquitous. Such systems include document databases (e.g., MongoDB),
or  systems combining a tree-structured 
data model and a columnar storage, such as F1~\cite{ShuteVSHWROLMECRSA13}, Dremel/BigQuery~\cite{DremelVLDBGLRSTVADMPS20,melnik2010dremel} and 
 Apache Parquet.


Identity and referential constraints (a.k.a., keys and foreign keys) have been extensively studied in the context of  relational databases and, lateron, in the context of XML data model~\cite{BunemanDFHT02,Fan05,FanS03}, as well as for graphs~\cite{FanL19} and RDF data~\cite{CalvaneseFPSS14,LausenMS08}. 
Recently, key constraints have been analyzed for JSON data models \cite{PezoaRSUV16,BourhisRSV17}. 

In this paper, we consider a theoretical \emph{tree-record data model}
for representing collections of tree-structured records.
This model is mainly inspired by the Dremel data model \cite{DremelVLDBGLRSTVADMPS20,melnik2010dremel,afrati2014dremel}, and it applies to document-oriented (i.e., XML, JSON) data stores (e.g., ElasticSearch, MongoDB) and relational databases supporting hierarchical data types (e.g., JSON type in PostgreSQL, MySQL and struct type in Hive). 
We define identity and referential constraints \emph{within each} tree-structured record (called \emph{within-record constraints}). 
Unlike relational databases and XML data, where such constraints are used for validating the data, in this work, we take advantage of them to improve query answering.
%
%
We consider SQL-like query language (such as the one used in Dremel,  F1,  Apache Drill),
and investigate querying in the presence of within-record constraints. 
We show that there are queries that can be answered only if we use the constraints.
To the best of our knowledge this is the first work investigating the problem of querying collections of tree-records with SQL  in the presence of such constraints.

There are many technical challenges that need to be addressed. 
The definition of keys is based on equality of values. In the relational model this is straightforward, but, in tree-structured data, we need first define precisely when we say that two instances of the schema are equal. We do that in a manner similar to such definitions for other tree-structured data such as XML, e.g.,  \cite{BunemanDFHT03}. We define the semantics of identity and reference constraints on the data by using flattening, and showing how to compute a query on the flattened data which  consists of tables like relational data. However, the traditional flattening is not able to handle keys that are referred to by more than one foreign key. This is a new challenge that we address for the first time in this paper. Another challenge lies in the definition of keys and foreign keys so that inconsistencies do not appear; we discuss this and point out to further research.

\emph{Contributions:}  (1) We define within-record key and foreign key constraints for the tree-structured data model (Section~\ref{sec:constraints}). 
(2) We show when these concepts are well-defined (Section~\ref{sec:safe-ids-refs}).
(3) We show how to use flattening to answer single-table, SQL-like queries (i.e., without joins) in the presence of such constraints
(Section~\ref{sec:querying}). We also introduce the concept of relative flattening which is a part of the flattened data corresponding to a subtree of the schema.
\section{Defining the data model}
\label{sec:data-model}


The tree-record data model considers collections of records (or, \emph{tables}) conforming to a predefined schema. 
We consider a nested, tree-structured schema, called \emph{tree-schema} (schema, for short) which uses 
primitive data types (such as integer, string, float, 
Boolean, etc.) to store the data and a complex data type, called \textit{group type},
to define the relationships between the data values and describe nested data structures. 

A \emph{group type} (or simply \textit{group}) $G$ is a complex data type defined by an ordered list of items (also called \emph{attributes} or \emph{fields}) of unique names which are associated with a data type, either primitive type or group type.
In fact, group type could be thought of as an element in XML, an object in JSON,
or as a Struct type in other data management systems (e.g., SparkSQL, Hive). 

We use a multiplicity constraint (also called \textit{repetition}) to specify the number of times a field is repeated within a group. Formally, the repetition constraint for a field $N$ can take one of the following values with the corresponding {\em annotation}:

{\squishlist
	\item \textit{required}: $N$ is mandatory, and there is no annotation,
	\item \textit{optional}: $N$ is optional (i.e., appears $0$ or $1$ times) and is labeled by $N?$,
	\item \textit{repeated}: $N$ appears $0$ or more times and is labeled by $N*$,
	\item \textit{required and repeated}: $N$  appears $1$ or more times and is labeled by $N+$.	
\squishend }
We denote as $repTypes$ the set $\{$required, repeated, optional, required and repeated$\}$ of repetition types. 
Note that a repeated field can be thought of as an array of elements (repeated types) in JSON structures. 
%
We now represent the tree-schema~\cite{afrati2014dremel} by a tree, as follows.

\begin{definition}\textbf{(tree-schema)} 
	A \emph{tree-schema} $S$ of a table $T$ is a tree with labeled nodes such that 
{\squishlist
		\item each non-leaf node (called \emph{intermediate node}) is a group and its children are its attributes, 
		\item each leaf node is associated with a primitive data type,
		\item each node (either intermediate or leaf) is associated with a repetition constraint in $repTypes$, and
		\item the root node is labeled by the name $T$ of the table.
\squishend }
\end{definition}

For the sake of simplicity, we hide the  primitive data types of leaves in the graphical representations of the schemas.  
Each non-required node is called \emph{annotated node}. When we de-annotate a node, we remove the repetition symbol from its label, if it is annotated.
$lb(N)$ represents the de-annotated label of a node $N$ in a schema. Considering the nodes $N_i$, $N_j$ of a schema $S$, such that $N_j$ is a descendant of $N_i$, we denote by $N_i\textbf{.}N_{i+1}\textbf{.}\dots\textbf{.}N_j$ the path between $N_i$ and $N_j$ in $S$. The path of de-annotated labels between the root and a node $N$ of a schema $S$ is called \textbf{reachability path} of $N$. 
%
%
We omit a prefix in the reachability path of a node if we can still identify the node through the remaining path.

Let us now define an instance of a tree-schema $S$. 
Considering a subtree $s$ of $S$, we denote as \emph{dummy} $s$ the tree constructed from $s$ by de-annotating all the annotated nodes of $s$ and adding to each leaf a single child which is labeled by the $NULL$-value. 

\begin{definition}\textbf{(tree-instance)} Let $S$ be a tree-schema and $t$ be a tree that is constructed from $S$ by recursively replacing, from top to down, each subtree $s_N$ rooted at an annotated node $N\sigma$, where $\sigma$ is an annotation, with
{\squishlist
		\item either a dummy $s_N$ or $k$ $s_N^d$-subtrees, if $\sigma=*$ \textbf{(repeated)},
		\item either a dummy $s_N$ or a single $s_N^d$-subtree, if $\sigma=?$ \textbf{(optional)},
		\item $k$ $s_N^d$-subtree, if $\sigma=+$ \textbf{(repeated and required)},
\squishend }
where $k\geq 1$ and $s_N^d$ is constructed from $s_N$ by de-annotating only its root.
Then, for each non-$NULL$ leaf $N$ of $t$, we add  to $N$ a single child which is labeled by a value of type that matches  the primitive type of  $N$. The tree $t$ is a \emph{tree-record} of $S$. 
An instance of $S$, called \emph{tree-instance}, is a multiset of tree-records. 
\end{definition}

\begin{example}
	\label{ex:tree-schema}
	Consider the table Booking with schema $S$ depicted in the Figure~\ref{fig:bookingschemaref}. At this stage, we ignore the $r$-labeled edges, which will be defined in the next section. The Booking table stores data related to reservations; each record in the table represents a single reservation. As we see in $S$, the Booking group includes a repeated and required $Service$ field (i.e., $Service+$) whose reachability path is $Booking.Service$; i.e., each booking-record includes one or more services booked by the customer. The $Type$ field describes the service type, is mandatory (i.e., required), and takes values from the set $\{$accommodation, transfers, excursions$\}$.  
	$Booking$ and $Service$ groups could include additional fields, such as date the reservation booked, start and end date of the service, that are ignored here due to space limitation. Figure~\ref{fig:bookinginst} illustrates a tree-record of $S$.
\end{example}

In the figures, all the dummy subtrees are ignored.
The reacha\-bility path of a node $N$ in a tree-record is similarly defined as the path from the root of the tree-record to $N$. 
We also consider that each node of both tree-schema and tree-record has a unique virtual id (called \emph{node id}).

In this paragraph, we define an instantiation in a multiset rather than in a set notion. Let $S$ be a schema and $t$ be a tree-record in an instance of $S$. 
Since each node of $S$ is replaced by one or more nodes in $t$, there is at least one mapping $\mu$, called \emph{instantiation}, from the node ids of $S$ to the node ids of $t$, such that ignoring the  annotations in $S$, both the de-annotated labels and the reachability paths
of the mapped nodes match. 
%
%
The subtree of $t$ which is rooted at the node $\mu(N)$ is an \emph{instance} of the node $N$, where $N$ is a node of $S$. If $N$ is a leaf, an instance of $N$ is the single-value child of $\mu(N)$.

We say that two subtrees $s_1$, $s_2$ of a tree-record are \emph{isomo\-rphic} if there is a bijective mapping $h$ from $s_1$ to $s_2$ such that the de-annotated labels of the mapped nodes match. We say that $s_1$ and $s_2$ of $t$ are equal, denoted $s_1=s_2$,  if they are isomorphic and the ids of the mapped nodes are equal.

\section{Within-record constraints}
\label{sec:constraints}
In this section, we define identity and referential constraints that hold on each tree-record.
%
Initially, we analyze the intuition behind the within-record constraints, and specifically, focus on identity and uniqueness constraints. In the conventional databases, we use identity (and uniqueness)  constraints (primary keys and unique constraints) to specify that the values of certain columns are unique across all the records of a table instance. Since tree-record data model allows repetition of values within each record, we might have fields that uniquely identify other fields (or subtrees) in each record, but not the record itself \cite{BunemanDFHT03}. 
In Example~\ref{ex:tree-schema}, each service has an identifier which is unique for each service within a reservation, but not unique across all the reservations in the Booking table.
To support such type of constraints in a tree-record data model, we define the concept of \textit{identity constraint with respect to a group}.

\begin{definition}\textbf{(identity constraint)}
	\label{den:identity-constraint}
		Let $S$ be a tree-schema with root $R_o$, $\CD$ be a tree-instance of $S$, and $N$, $I$ be nodes of $S$ such that $N$ is intermediate and $I$ is a descendant of $N$. Suppose that $M$ is the parent of $N$, if $N\neq R_o$; otherwise, $M=N=R_o$.
	An \emph{identity constraint with respect to} $N$ is an expression of the form $I\idarr N$, such that $I$ and all the descendants of $I$ in $S$ are required. 
	We say that $I\idarr N$ \emph{is satisfied} in $\CD$ if for each $t\in\CD$ and for each instance $t_p$ of $M$  in $t$, there are not two isomorphic instances of $I$ in $t_p$.
	The node $I$ is called \emph{identifier} and $N$ is the \emph{range group} of $I$. 
\end{definition}

The identity constraint is similar to the concept of relative key defined for XML documents in \cite{BunemanDFHT03}. Intuitively, if the constraint $I\idarr N$ is satisfied in $\CD$ and there is a repeated node between $N$ and $I$ then for each tree-record $\CD$, the instances of $I$ uniquely identify the instances of the lowest repeated ancestor of $I$ within each instance of the node $N$.
%
If there is no repeated node between $N$ and $I$ then there is a single, unique instance of $I$ in each instance of $N$.
If $I\idarr R_o$, where $R_o$ is the root of $S$, then all the instances of $I$ are unique in each tree-record.

In the tree representation of a schema, for each $I\idarr N$, we use the symbol $\#$ to annotate the identifier $I$ (i.e., $I\#$). We also use a special, dotted edge $(I,N)$, called \emph{identity edge}, to illustrate the range group $N$ of $I$. If $N$ is the parent of $I$, we omit such an edge, for simplicity.

\begin{figure}[t!]
	\centering
	\includegraphics[width=0.7\linewidth]{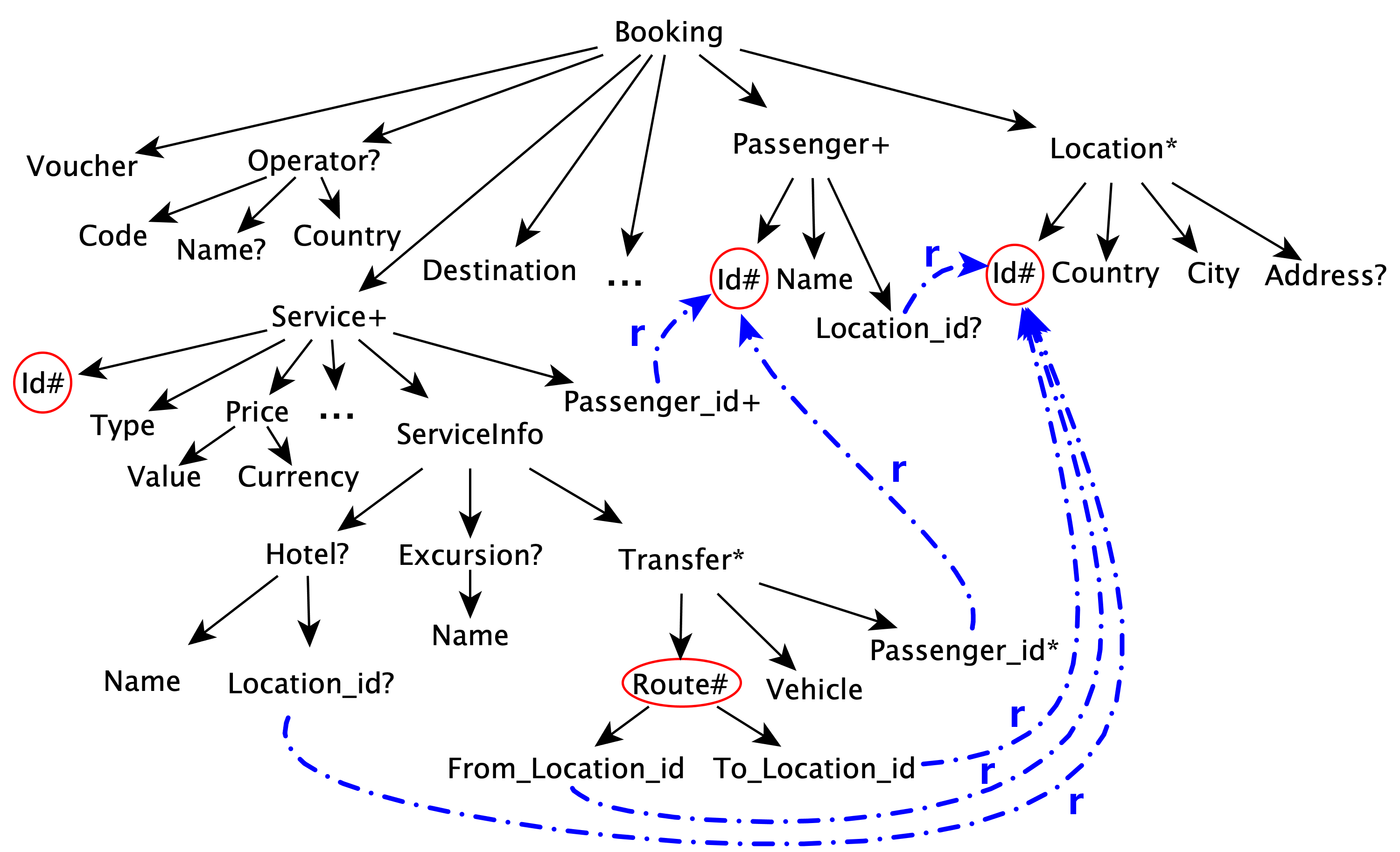}
	\caption{Booking Schema - Tree-record model with references}
	\label{fig:bookingschemaref}
\end{figure}

\begin{figure*}
	\centering
	\includegraphics[width=1.0\linewidth]{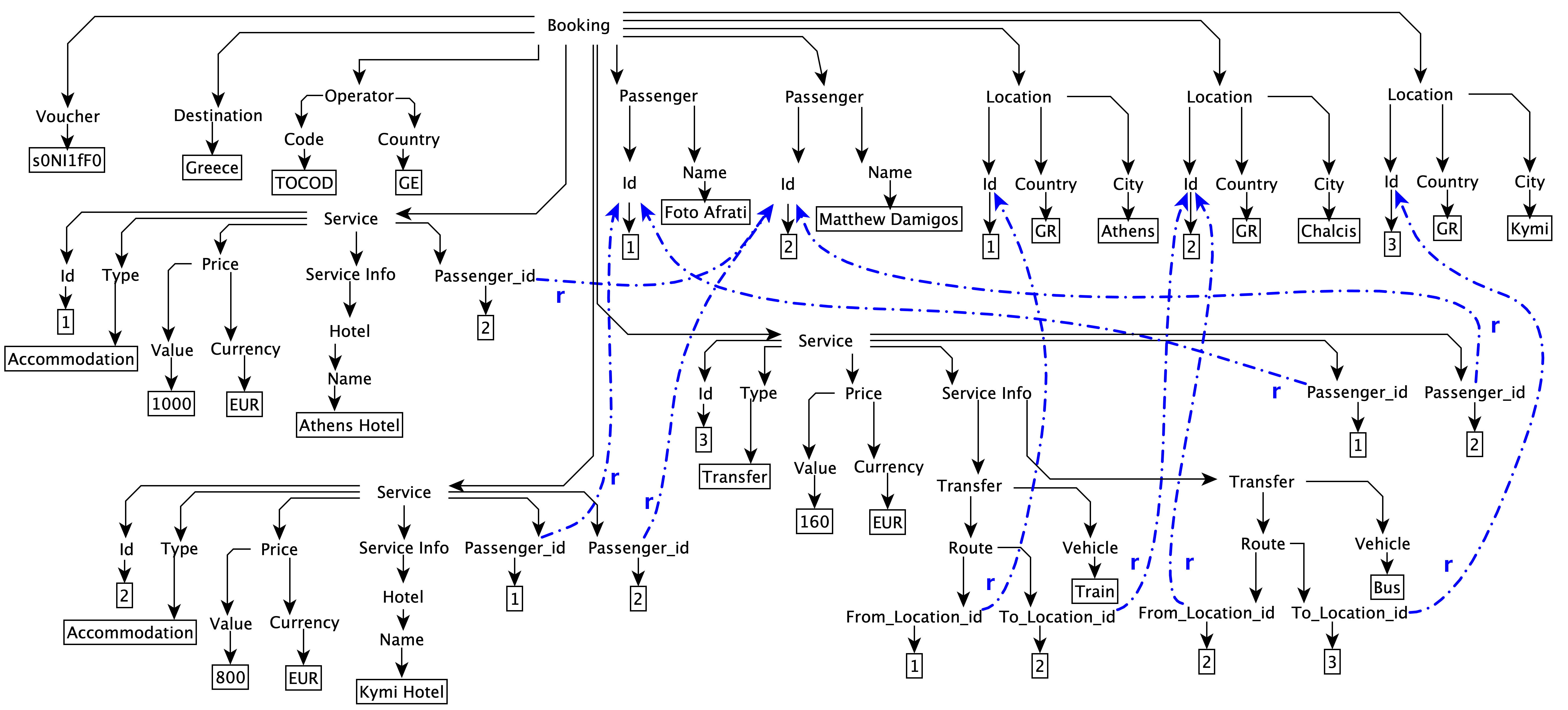}
	\caption{Booking instance - Tree-record with references}
	\label{fig:bookinginst}
\end{figure*}

\begin{example}
	\label{ex:tree-schema-ids}
	Continuing the Example~\ref{ex:tree-schema}, each reservation-record includes a list of passengers which is given by the field $Passenger+$. The $Passenger$ includes 3 fields $Passenger.Id$, $Passenger.Name$, $Passenger.Loca\-tion\_id$,  
	where the last one is optional for each passenger. Notice also that there is the identity constraint $Passenger.Id\idarr Passenger$, which means that the field $Passenger.Id$ uniquely identifies the $Passenger.$ Since the range group of the $Passenger.Id$ is its parent, we ignore the corresponding identity edge. In Figure~\ref{fig:bookinginst}, we can see a tree-record that satisfies this constraint, since each $Passenger$ instance has a unique $Id$.   
	
	To see the impact of the range group, let us compare the following two constraints:  $Route\idarr Transfer$ and  $Route\idarr Service$.
	The field $Route$ in the former case (i.e., the one illustrated in Figure~\ref{fig:bookingschemaref}) is a composite identifier of its parent and consists of two location ids; i.e., the combination of From and To locations uniquely identifies the transfer instances within each service, but not across all the services of the booking. On the other hand, setting the range group of the $Route$ to $Service$ (i.e., the latter constraint), the combination of From and To locations are unique across all the transfer services in each booking-record.
\end{example}

We now define the concept of \textit{referential constraint} (or, simply \emph{reference}), which intuitively links the values of two fields.
In essence, the concept of reference is similar to the foreign key in relational databases, but, here, it is applied within each record.

\begin{definition}\textbf{(referential constraint)}
	\label{den:reference-constraint}
	Let $I$, $N$ and $R$ be nodes of a tree-schema $S$ such that $I\idarr N$, $R$ is not a descendant of $N$, and $I$, $R$ have the same data type. A \emph{referential constraint} is an expression of the form $R\refarr I$. A tree-instance $\CD$ of $S$ \emph{satisfies} the constraint $R\refarr I$, if for each tree-record $t\in\CD$ the following is true:
	For each instance $t_{LCA}$ of the lowest common ancestor (LCA) of $R$ and $N$ in  $t$, each instance of $R$ in $t$ is isomorphic to an instance of $I$ in $t_{LCA}$. If $I$ is a leaf, then $R\refarr I$ is called \emph{simple}. 
\end{definition}

If we have $R\refarr I$, we say that $R$ (called \emph{referrer}) \emph{refers} to $I$ (called \emph{referent}). 
To represent the constraint $R\refarr I$ in a tree-schema $S$, we add a special (dashed) edge $(I,R)$, called \emph{reference edge}, which is labeled by $r$. Let $\CC$  be the set of identity and referential constraints over $S$. Consider now the tree $B$ given by (1) ignoring all the reference and identity edges, and (2) de-annotating the identifier nodes. 
%
We say that a collection $\CD$ is a tree-instance of the tree-schema $S$  in the presence of $\CC$ if $\CD$ is a tree-instance of $B$ and $\CD$ satisfies all the constraints in $\CC$.


\begin{example}
	\label{ex:tree-schema-refs}
	Continuing the Example~\ref{ex:tree-schema-ids}, we can see that the schema $S$ depicted in Figure~\ref{fig:bookingschemaref} includes two referrers of the $Passenger.Id$; $Service.Passenger\_Id$ and $Transfer.Passenger\_Id$ store the ids of the passengers that booked each service and the ids of the passe\-ngers taking each transfer, respectively.
	Further\-more, the $Location.Id$ is a referent in four references defined, while the composite identifier $Route$ consists of two fields that both refer to the $Location.Id$ identifier.
	
	
	Consider now the tree-record $t$ of $S$ which is depicted in Figure~\ref{fig:bookinginst}. It is easy to see that this reservation includes 3 services booked for two passengers. 
	The first service is booked for the first passenger, while the services with ids 2 and 3 are taken by both passengers. In each service, there is a list of passenger ids representing the passengers taking each service. Those fields refer to the corresponding passengers in the passenger list of the booking; appropriate reference edges illustrate the references in $t$.
	
\end{example}

\section{Querying tree-structured data}
\label{sec:querying}
In this section, we investigate querying tree-structured tables in the presence of identity and referential constraints. Although navigation languages (e.g., XPath,
XQuery,
JSONPath~\footnote{JSONPath (2007). http://goessner.net/articles/JsonPath})
are used to query a single tree-structured document, here, we focus on a combination of a simple navigation language and SQL-like language to query collections of tree-structured data.
In particular, we use the  Select-From-Where-GroupBy expressions\footnote{In this work, we do not consider joins, recursion, nested queries and within-aggregations~\cite{melnik2010dremel}, as well as operations that are used to build a tree-like structure at query-time, or as a result of the query (e.g., the $json\_build$-like functions in PostgreSQL).} used in Dremel~\cite{melnik2010dremel,afrati2014dremel} to query tables defined through a tree-schema. 
We refer to such a query language as \emph{Tree-SQL}.

A query $Q$ is an expression over a table $\CT$ with schema $S$ and tree-instance 
$\CD$ in the presence of within-constraints $\CC$, and results a relation (multiset of tuples). 
We consider \emph{only} simple references in $\CC$~\footnote{Querying schemas having references to intermediate nodes is considered a topic for future investigation.}.
The expression $Q$ has the following form, using the conventional SQL syntax~\cite{GMUWBookDB2nd}:\\
$\textbf{SELECT}\; expr\; \textbf{FROM}\; \CT\; [\textbf{WHERE}\; cond]\; [\textbf{GROUP}\; \textbf{BY}\; grp],$\\
where $cond$ is a logical formula over fields of $S$ and $grp$ is a list of grouping fields. $cond$, $expr$ and $grp$ are defined in terms of the leaves of $S$. $expr$ is a list of selected leaves of $S$ followed by a list of aggregations over the leaves of $S$. In the case that $expr$ includes both aggregated operators and fields that are not used by aggregations, those fields should be present in the $GROUP$ $BY$ clause.  Each leaf node in $Q$ is referred through either its reachability path or a path using reference and identity edges implied by the constraints in $\CC$.

A query typically determines a mapping from tree-structured data model to relational model; i.e., it unnests the tree-structured data and transforms the tree-records into tuples. To analyze the semantics of a query in more detail, we initially ignore the references and identifiers. Consider, for example, the following query $Q$ over the table $Booking$ with schema $S$ depicted in Figure~\ref{fig:bookingschemaref}:
\begin{center}
	\small
	\begin{verbatim}
		SELECT Voucher, Destination, Operator.Name
		FROM Booking
		WHERE Operator.Country='GE';
	\end{verbatim}
\end{center}
%
When $Q$ is applied on an instance $\CD$ of $S$ it results a relation, also denoted $Q(\CD)$, including a single tuple for each tree-record $t$ in $\CD$ such that the instance of the $Operator.Country$ field in $t$ is \verb:'GE':. Each tuple in $Q(\CD)$ includes the voucher of the booking, the destination and the name of the operator (if it exists - otherwise, the $NULL$-value). For example, if $\CD$ includes the tree-record illustrated in Figure~\ref{fig:bookinginst}, then $Q(\CD)$ includes the tuple ($s0NI1fF0$, $Greece$, $NULL$).

In the previous example, we can see that the fields used in both $SELECT$ and $WHERE$ clauses do not have any repeated field in their reachability path. Querying the instances of such kind of fields is similar to querying a relation consisting of a column for each field. The tuples are constructed by assigning the single value of each field, in each record, to the corresponding column.  The challenge comes up when a repeated node exists in the reachability path of a field used in the query; since such a field might have multiple instances in each tree-record. To formally define the query semantics and handle repetition, we use the concept of flattening~\cite{afrati2014dremel} which is discussed in detail in the next section.
\vspace{-3mm}
\subsection{Flattening nested data}

In this section, we analyze the flattening operation applied on tree-structured data. Flattening is a mapping applied on a tree-structured table and translates the tree-records of the table to tuples in a relation. By defining  such a mapping, the semantics of Tree-SQL is given by the conventional SQL semantics over the \emph{flattened relation} (i.e., the result of the flattening over the table). Initially, we consider a tree-schema without referential and identity constraints. The presence of constraints is discussed in the next section.

Let $S$ be a tree-schema of a table $T$ and $\CD$ is an instance of $S$, such that there is not any reference defined in $S$. Suppose also that $N_1,\dots , N_m$ are the leaves of $S$. 
The \emph{flattened relation} of $\CD$, denoted $flatten(\CD)$, is the relation given by the multiset:
$\{\{(lb(\mu(N_1)),\dots, lb(\mu(N_m)))\;|\;\mu$ is an instantiation of a tuple $t\in\CD\}\}$.
For each pair $(N_i,N_j)$, $\mu(N_i)$ and $\mu(N_j)$ belong to the same instance of the lowest common ancestor of  $N_i$ and $N_j$ in $t$. Considering now a query $Q$ over $S$ and an instance $\CD$ of $S$, we say that $Q$ is evaluated using \emph{full flattening}, denoted $Q(flatten(\CD))$,  if $Q(\CD)$ is given by evaluating $Q$ over the relation $flatten(\CD)$. It's worth noting here that if $S$ does not have any repeated field then the flattened relation of $\CD$ includes $|\CD|$ tuples; otherwise, each record in $\CD$ can produce multiple tuples during flattening.

\begin{figure}[h]
	\centering
	\includegraphics[width=0.80\linewidth]{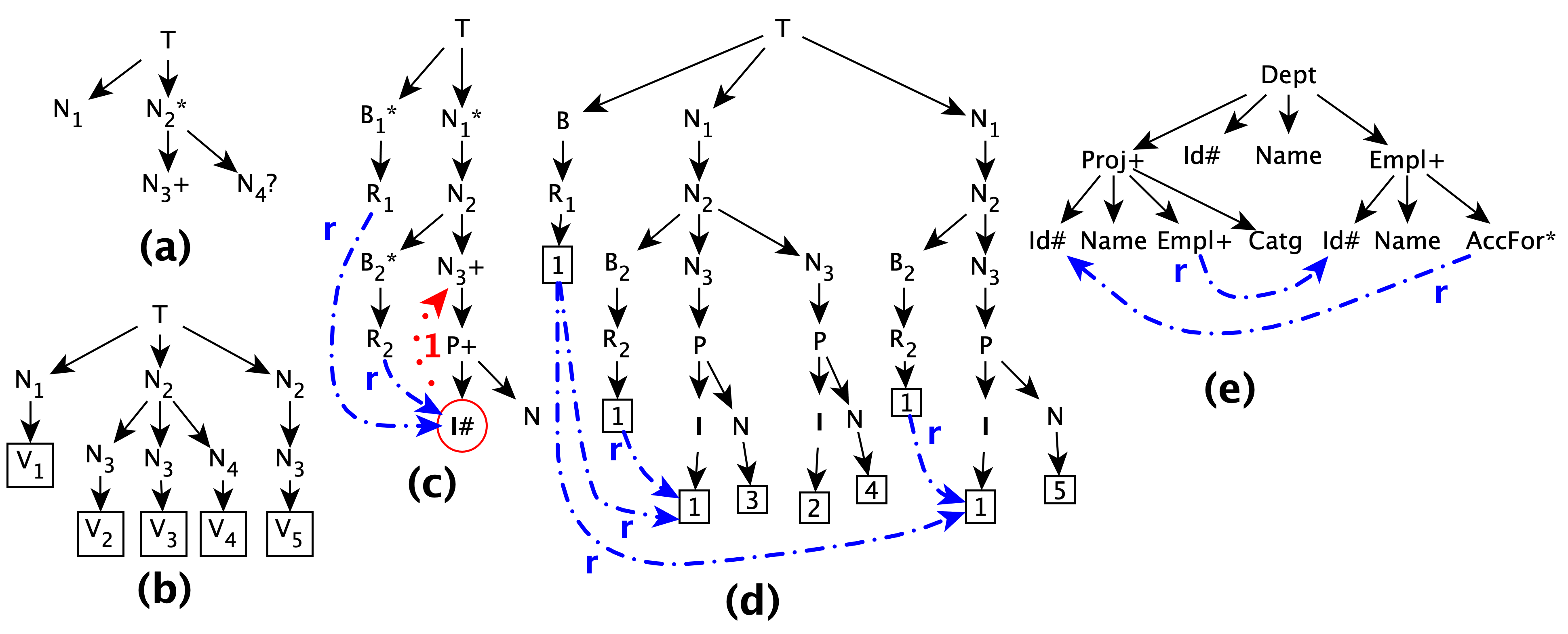}
	\caption{Flattening, out-of-range references and cycles}
	\label{fig:flat-special-cases}
\end{figure}


\begin{example}
	\label{ex:flattening}
	Let $T$  be a table with schema $S$ depicted in Figures~\ref{fig:flat-special-cases}(a), and instance $\CD$ including only the tree-record depicted in Figures~\ref{fig:flat-special-cases}(b). 
	The flattened relation $flatten(\CD)$ is $\{\{(V_1,V_2,V_4)$, $(V_1,V_3,V_4)$, $(V_1,V_5,NULL)\}\}$.
	Consider now the query $Q$: 
	$SELECT$ $N_1$, $N_4$ $FROM$ $T$. $Q$ typically applies a projection over the flattened relation; hence, it results three tuples; i.e., $\{$($V_1$, $V_4$), ($V_1$, $V_4$), ($V_1$, $NULL$)$\}$. However, we see that $N_4$ in $\CD$ has two instances, $V_4$ and $NULL$. The evaluation of $Q$ using full flattening is affected by the repetition of $N_3$. 
\end{example}

Now we motivate the definition of relative flattening with an example. Suppo\-sing the table $\CT$ with schema and record depicted in Figures~\ref{fig:bookingschemaref} and~\ref{fig:bookinginst}, respectively, we want to find the total price 
(i.e., $sum(Service.Price)$) 
for accommodation services.
Although the total price is $1,800$, we can see that using full flattening, the result of the query is $9,000$; due to the repetition of the $Passenger$ and $Location$ subtrees.

To avoid cases where the repetition of a field that is not used in the query has an impact on the query result, we define the concept of \emph{relative flattening}. Let $S$ be a tree-schema of a table $T$ and $\CD$ is an instance of $S$, such that there is not any reference defined in $S$. Consider also a query $Q$ over $T$ that uses a subset  $L=\{N_1,\dots,N_k\}$ of the set of leaves of $S$, and the tree-schema $S_L$ constructed from $S$ by removing all the nodes except the ones included in the reachability paths of the leaves in $L$. Then, we say that a query $Q$ is evaluated using \emph{relative flattening}, denoted $Q(flatten(\CD,Q))$, if $Q(\CD)$ is given by evaluating $Q$ over the relation:
$flatten(\CD,Q)=\{\{(lb(\mu(N_1)),\dots, lb(\mu(N_k)))\;|\;\mu$ is an instantiation from the nodes of $S_L$ to the nodes of $t\in\CD\}\}$.
Continuing the Example~\ref{ex:flattening}, we have that $Q(flatten(\CD,Q))=$ $\{\{(V_1,V_4)$, $(V_1,NULL)\}\}$.
\begin{proposition}
	Consider a query $Q$ over a tree-schema $S$ such that $Q$ does not apply any aggregation. Then, for every instance $\CD$ of $S$, the following are true:\\
	(1) there is a tuple $r$ in $Q(flatten(\CD,Q))$ if and only if there is a tuple $r$ in $Q(flatten(\CD))$,\\
	(2) $|Q(flatten(\CD,Q))|\leq |Q(flatten(\CD))|$.
\end{proposition}	

%
%
%
\vspace{-3mm}

\subsection{Navigating through references}
In the previous section, we ignored the presence of constraints when we explained how to use flattening to answer an SQL-like query. Here, we show how we take advantage of the constraints to extend the query semantics based on the relative flattening.

Let us start our analysis by looking at the schema $S$ in Figure~\ref{fig:bookingschemaref}. Let $\CD$ be an instance of $S$. Suppose now that we want to find, for all the transfer services in $\CD$, their vouchers, along with  the following transfer information: vehicle of each transfer and the route expressed as a combination of From and To cities. Note that this query cannot be answered based on the query semantics defined in the previous section\footnote{If the data is structured as in Figure~\ref{fig:bookingschemaref}, such a query cannot be answered using Select-From-Where-GroupBy queries in Dremel, as well.}, since the city of each location does not belong into the same Route subtree. Taking into account, however, the following constraints, it is easy to see that intuitively such a query could be answered.\\
\begin{tabular}{rcl}
	$From\_Location\_id\refarr Location.Id$ & $\;\;$ & $Location.Id\idarr Location$\\
	$To\_Location\_id\refarr Location.Id$ & & \\
\end{tabular}\\
To see this, we can initially search for the voucher, vehicle, and ids of the From and To Locations for each transfer service within all the bookings. Then, for each id of the From and To Locations, we look at the corresponding Location list of the same record and identify the corresponding cities. To capture such cases and use the identity and referential constraints, we initially extend the notation of the Tree-SQL as follows. Apart from the reachability paths of the leaves that can be used in $SELECT$, $WHERE$ and $GROUP\; BY$ clauses, if there are constraints $R\refarr I$ and $I\idarr G$, we can use paths of the form:
$[pathToR].R.[pathToL]$,
where the $[pathToR]$ is the reachability path of $R$, $L$ is a leaf which is a descendant of $G$, and $[pathToL]$ is the path from $G$ to $L$. Hence, the query $Q_{tr}$ answering the previous question:\\
{\small\verb|SELECT	Voucher, Vehicle, Route.From_Location_id.City, |}\\
{\small\verb|Route.To_Location_id.City |}\\
{\small\verb|FROM Booking WHERE Service.Type = 'transfer';|}\\
Intuitively, navigating through identity and reference edges, the leaves of $G$ become accessible from $R$. For example, in the schema $S$ in Figure~\ref{fig:bookingschemaref}, the leaves of the group $Location$ are accessible through both $From\_Location\_id$ and $To\_Location\_id$.

To formally capture queries using references, we extend the relative flattening presented in the previous section as follows. 
Let $S$ be a tree-schema of a table $T$, $\CD$ be an instance of $S$, and $\CC$ be a set of identity and referential constraints satisfied in $\CD$. Consider also a query $Q$ over $T$ that uses a set of leaves  $\CL=\{N_1,\dots,N_k, L_1,\dots,L_m\}$ of $S$ such that for each $L_i$ in $\CL$ there are constraints $R_i\refarr I_i$, $I_i\idarr G_i$ in $\CC$, where $G_i$ is an ancestor of $L_i$. Let also $S_{\CL}$ be the tree-schema constructed from $S$ by keeping only the reachability paths of the leaves in $\{N_1,\dots,N_k, R_1,\dots,R_m\}$ (without de-annotating any node), and for each $i$, $S_{G_i}$ be the tree-schema including only the reachability paths of $R_i$, $I_i$ and the leaves of $G_i$ that are included in $\{L_1,\dots,L_m\}$. Both $S_{\CL}$  and $S_{G_i}$ keep the node ids from $S$. A query $Q$ is evaluated in the presence of the constraints in $\CC$, denoted $Q(flatten(\CD,Q,\CC))$, if $Q(\CD)$ is given by evaluating $Q$ over the relation:
$flatten(\CD,Q,\CC)$ $=$ $\{\{($$lb(\mu(N_1)),$ $\dots,$ $lb(\mu(N_k)),$ $lb(\mu_1(L_1)),$ $\dots,$ $lb(\mu_m(L_m)))$ $|\;\mu$ is an instantiation from the nodes of $S_{\CL}$ to the nodes of $t\in\CD$, each $\mu_i$ is an instantiation from the nodes of $S_{G_i}$ to the nodes of $t$, for every two $L_i$, $L_j$ s.t. $G_i=G_j$ and $R_i=R_j$, we have that $\mu_i=\mu_j$, and for each $i$, we have that $\mu(R_i)=\mu_i(R_i)$ and $lb(\mu_i(R_i))=lb(\mu_i(I_i))\}\}$. 

Posing $Q_{tr}$ (defined above) on an instance $\CD$ including the tree-record depicted in Figure~\ref{fig:bookinginst}, we have two instantiations, each of which maps on a different instance of $Transfer$ subtree. For each such instantiation, there is a single instantiation to a $Location$ instance such that the referrer value equals the $Location.Id$ value. The result $Q_{tr}(\CD)$ is: $\{(s0NI1fF0,$ $Train,$ $Athens,$ $Chalcis)$, $(s0NI1fF0,$ $Bus,$ $Chalcis,$ $Kymi)\}$.
If we replace 
$Route.$$To\_Location\_id.$$City$ with the field $Location.City$, the refe\-rence $To\_Location\_id\refarr Location.Id$ is not used; hence, the result includes 6 tuples computed by combina\-tion of the 2 cities of From-location, $Athens$ and $Chalcis$, and all the available instances of the $Location.City$.
\vspace{-3mm}

\subsection{Out-of-range references and cycles}
\label{sec:safe-ids-refs}
In this section, we investigate well-defined references; i.e., whether it is clear which referent is referred by each referrer in a tree-instance. We also discuss cases where the references define a cycle into the schema graph. 


Consider the tree-schema $S$ depicted in Figure~\ref{fig:flat-special-cases}(c) and an instance $\CD$ of $S$ including the tree-record in Figure~\ref{fig:flat-special-cases}(d). As we can see, there are 2 referrers, $R_1$ and $R_2$, which both refer to the identifier $I$. The range group of $I$ is the node $N_3$. Consider the queries $Q_1$ and $Q_2$ selecting only the  fields $R_1$ and $R_2$, respectively. Note that the result $Q_2(\CD)$ includes two times the value $1$, while $Q_1(\CD)$ includes the value $1$ once. Let now $Q_1'$ and $Q_2'$ be the queries selecting the paths $R_1.N$ and $R_2.N$, respectively. We can see that both $Q_1'(\CD)$ and $Q_2'(\CD)$ include the tuples $(3)$ and $(5)$. Hence, when we use the reference from $R_2$, the number of tuples in the result remains the same. Using however the reference from $R_1$, the number of tuples in the result increases. This is because it is not clear which is the instance of $I$ that the instance of $R_1$ refers to. This property is captured by the following definition and proposition.
\vspace{-2mm}
\begin{definition}
	Let $S$ be a tree-schema, and $R\refarr I$, $I\idarr G$ be two constraints over $S$. Let $L$ be the lowest common ancestor of $G$ and $R$. Then, we say that the reference $R\refarr I$ is \emph{out-of-range} if there is at least one repeated group on the path from $L$ to $G$; otherwise, the reference is \emph{within-range}.
\end{definition}

\begin{proposition}
	\label{prop:out-of-range}
	Let $C$ be a reference $R\refarr I$ over a tree-schema $S$, and $t$ be a record of a tree-instance $\CD$ of $S$ satisfying the constraint. If $C$ is within-range, then for each instance of $R$ in $t$, there is a single instance of $I$ in $t$.
\end{proposition}
\vspace{-2mm}

The property described in the Proposition~\ref{prop:out-of-range} is very important for defining referential constraints, since setting up out-of-range constraints the queries using the references might not compute the "expected" results. 

By defining references in a tree-schema, cycles of references can appear. For example, consider the table $T$ with tree-schema $S$ illustrated in Figure~\ref{fig:flat-special-cases}(e).
The table $T$ stores the projects ($Proj$) of each department ($Dept$), along with the employees ($Empl$) of the department. Each project has a number of employees working on it, and each employee of the department might be accountable for ($AccFor$) a list of projects. Hence, it is easy to see that the references defined between $Proj$ and $Empl$ subtrees form a cycle of references. One could ask for the projects of a certain category ($Catg$) which employ an employee who is accountable for a project of a different category, along with the name of the employee. To answer such a question, we need to navigate through both links. The query semantics defined in the previous section allow only a single use of a reference between two subtrees. Extending the semantics to support arbitrary naviga\-tion through the references is a topic for future work.

%
%
%
%

\section{Related Work}

Work on constraints for tree-structured data has been done during the past two decades. 
Our work, as regards the formalism,  is closer to  \cite{VoCR11,YuJ08,BunemanDFHT03,BunemanDFHT02}. The  papers \cite{BunemanDFHT03} and \cite{BunemanDFHT02} are among the first works on defining constraints on tree-structured data. Reasoning about keys for XML is done in \cite{BunemanDFHT03} where a single document XML is considered and keys within scope (relative keys) are introduced. Referential constraints through inclusion dependencies are also investigated (via path expression containment). The satisfiability problem is investigated, but no query language is considered. 
Many recent works investigate discovering conditional functional dependencies in XML Data; closer to our perspective is \cite{VoCR11}   and  \cite{YuJ08} where XML schema refinement is studied through redundancy detection and normalization.

\cite{PezoaRSUV16} and \cite{BourhisRSV17} focus on the JSON data model and a similar to XPath navigational query language. These works also formalize specification of unique fields and references,  they do not define relative keys.
\cite{PezoaRSUV16} formally defines a JSON schema. It supports specification of unique fields within an object/element and supports references to an another subschema (same subschema can be used in several parts of the schema). No relative keys are supported.  \cite{BourhisRSV17} continues on \cite{PezoaRSUV16} and 
%
%
proposes a navigational query language over a single JSON document (this language presents XPath-like alternatives for JSON documents, such as JSONPath, MongoDB navigation expressions and JSONiq). 

Flattening has initially been studied in the context of nested relations and hierarchical model (e.g., \cite{scholl1987supporting,Colby89,paredaens1988possibilities}).
Dremel~\cite{melnik2010dremel,afrati2014dremel}, F1~\cite{ShuteVSHWROLMECRSA13} and Drill
use flattening to answer SQL-like queries over tree-structured data. Flattening semi-structured data is also investigated in \cite{DiScalaA16,LiuHM14,DeutschFS99}, where the main problem is to translate semi-structured data into multiple relational tables.


\section{Future work}
As next steps, we plan to investigate querying tree-schemas having references to intermediate nodes and/or reference cycles. Also, we aim to study flattening when the referrer is defined in the range group of the referent.  
Furthermore, we plan to extend this investigation towards the following directions: a) Study the satisfiability and the  implication problems for the constraints we defined here. b) The chase \cite{SadriU80} is used to reason about keys and functional dependencies. For relational data, there is a lot of work on chase. The chase for RDF and graph data was studied in \cite{HellingsGPW16}, \cite{FanFTD15,FanWX16a}, \cite{Corte-CalabuigP12} and \cite{FanL19}. We plan to define a new chase that can be applied to reason about the constraints we defined here. 
%
%
%
%
\bibliographystyle{splncs04}
\bibliography{bibliography}

\begin{thebibliography}{10}
\providecommand{\url}[1]{\texttt{#1}}
\providecommand{\urlprefix}{URL }
\providecommand{\doi}[1]{https://doi.org/#1}

\bibitem{afrati2014dremel}
Afrati, F.N., Delorey, D., Pasumansky, M., Ullman, J.D.: Storing and querying
  tree-structured records in dremel. Proc. {VLDB} Endow.  \textbf{7}(12),
  1131--1142 (2014)

\bibitem{BourhisRSV17}
Bourhis, P., Reutter, J.L., Su{\'{a}}rez, F., Vrgoc, D.: {JSON:} data model,
  query languages and schema specification. In: {PODS} 2017. pp. 123--135.
  {ACM} (2017)

\bibitem{BunemanDFHT02}
Buneman, P., Davidson, S.B., Fan, W., Hara, C.S., Tan, W.C.: Keys for {XML}.
  Comput. Networks  \textbf{39}(5),  473--487 (2002)

\bibitem{BunemanDFHT03}
Buneman, P., Davidson, S.B., Fan, W., Hara, C.S., Tan, W.C.: Reasoning about
  keys for {XML}. Inf. Syst.  \textbf{28}(8),  1037--1063 (2003)

\bibitem{CalvaneseFPSS14}
Calvanese, D., Fischl, W., Pichler, R., Sallinger, E., Simkus, M.: Capturing
  relational schemas and functional dependencies in {RDFS}. In: {AAAI} 2014.
  pp. 1003--1011. {AAAI} Press (2014)

\bibitem{Colby89}
Colby, L.S.: A recursive algebra and query optimization for nested relations.
  In: {SIGMOD} 1989. pp. 273--283 (1989)

\bibitem{Corte-CalabuigP12}
Cort{\'{e}}s{-}Calabuig, A., Paredaens, J.: Semantics of constraints in {RDFS}.
  In: {AMW} 2012. {CEUR} Workshop Proceedings, vol.~866, pp. 75--90.
  CEUR-WS.org (2012)

\bibitem{DeutschFS99}
Deutsch, A., Fern{\'{a}}ndez, M.F., Suciu, D.: Storing semistructured data with
  {STORED}. In: {SIGMOD} 1999. pp. 431--442 (1999)

\bibitem{DiScalaA16}
DiScala, M., Abadi, D.J.: Automatic generation of normalized relational schemas
  from nested key-value data. In: {SIGMOD} 2016. pp. 295--310 (2016)

\bibitem{Fan05}
Fan, W.: {XML} constraints: Specification, analysis, and applications. In: 16th
  International Workshop on Database and Expert Systems Applications {(DEXA}
  2005). pp. 805--809. {IEEE} Computer Society (2005)

\bibitem{FanFTD15}
Fan, W., Fan, Z., Tian, C., Dong, X.L.: Keys for graphs. Proc. {VLDB} Endow.
  \textbf{8}(12),  1590--1601 (2015)

\bibitem{FanL19}
Fan, W., Lu, P.: Dependencies for graphs. {ACM} Trans. Database Syst.
  \textbf{44}(2),  5:1--5:40 (2019)

\bibitem{FanS03}
Fan, W., Sim{\'{e}}on, J.: Integrity constraints for {XML}. J. Comput. Syst.
  Sci.  \textbf{66}(1),  254--291 (2003)

\bibitem{FanWX16a}
Fan, W., Wu, Y., Xu, J.: Functional dependencies for graphs. In: {SIGMOD}
  Conference 2016. pp. 1843--1857. {ACM} (2016)

\bibitem{GMUWBookDB2nd}
Garcia{-}Molina, H., Ullman, J.D., Widom, J.: Database systems - the complete
  book {(2.} ed.). Pearson Education (2009)

\bibitem{HellingsGPW16}
Hellings, J., Gyssens, M., Paredaens, J., Wu, Y.: Implication and
  axiomatization of functional and constant constraints. Ann. Math. Artif.
  Intell.  \textbf{76}(3-4),  251--279 (2016)

\bibitem{LausenMS08}
Lausen, G., Meier, M., Schmidt, M.: Sparqling constraints for {RDF}. In: {EDBT}
  2008. vol.~261, pp. 499--509. {ACM} (2008)

\bibitem{LiuHM14}
Liu, Z.H., Hammerschmidt, B.C., Mcmahon, D.: {JSON} data management: supporting
  schema-less development in {RDBMS}. In: {SIGMOD} 2014. pp. 1247--1258 (2014)

\bibitem{melnik2010dremel}
Melnik, S., Gubarev, A., Long, J.J., Romer, G., Shivakumar, S., Tolton, M.,
  Vassilakis, T.: Dremel: interactive analysis of web-scale datasets. Proc.
  {VLDB} Endow.  \textbf{3}(1-2),  330--339 (2010)

\bibitem{DremelVLDBGLRSTVADMPS20}
Melnik, S., Gubarev, A., Long, J.J., Romer, G., Shivakumar, S., Tolton, M.,
  Vassilakis, T., Ahmadi, H., Delorey, D., Min, S., Pasumansky, M., Shute, J.:
  Dremel: {A} decade of interactive {SQL} analysis at web scale. Proc. {VLDB}
  Endow.  (2020)

\bibitem{paredaens1988possibilities}
Paredaens, J., Van~Gucht, D.: Possibilities and limitations of using flat
  operators in nested algebra expressions. In: PODS 1988. pp. 29--38 (1988)

\bibitem{PezoaRSUV16}
Pezoa, F., Reutter, J.L., Su{\'{a}}rez, F., Ugarte, M., Vrgoc, D.: Foundations
  of {JSON} schema. In: {WWW} 2016. pp. 263--273 (2016)

\bibitem{SadriU80}
Sadri, F., Ullman, J.D.: The interaction between functional dependencies and
  template dependencies. In: {ACM} {SIGMOD} Conference 1980. pp. 45--51. {ACM}
  Press (1980)

\bibitem{scholl1987supporting}
Scholl, M.H., Paul, H.B., Schek, H.J., et~al.: Supporting flat relations by a
  nested relational kernel. In: VLDB. pp. 137--146 (1987)

\bibitem{ShuteVSHWROLMECRSA13}
Shute, J., Vingralek, R., Samwel, B., Handy, B., Whipkey, C., Rollins, E.,
  Oancea, M., Littlefield, K., Menestrina, D., Ellner, S., Cieslewicz, J., Rae,
  I., Stancescu, T., Apte, H.: {F1:} {A} distributed {SQL} database that
  scales. Proc. {VLDB} Endow.  \textbf{6}(11),  1068--1079 (2013)

\bibitem{VoCR11}
Vo, L.T.H., Cao, J., Rahayu, J.W.: Discovering conditional functional
  dependencies in {XML} data. In: {ADC} 2011. {CRPIT}, vol.~115, pp. 143--152
  (2011)

\bibitem{YuJ08}
Yu, C., Jagadish, H.V.: {XML} schema refinement through redundancy detection
  and normalization. {VLDB} J.  \textbf{17}(2),  203--223 (2008)

\end{thebibliography}

%
% ...

\end{document}